# Broadband Simultaneous Beam Steering and Compressing Device Based on Subwavelength Protrusion Metallic Tunnels

Dongguo Zhang, Fei Sun, Qin Liao, Yichao Liu, and Donguk Nam

*Abstract*—Beam steering and beamwidth compressing play a role in steering the beam and narrowing its half-power beamwidth, respectively, which are both widely applied in extending the effective operational range of 6G communications, IoT devices, and antenna systems. However, research on wave manipulation devices capable of simultaneously achieving both functionalities remains limited, despite their great potential for system miniaturization and functional integration. In this study, we design and realize a broadband device capable of simultaneously steering and compressing the TM-polarized EM waves using subwavelength protrusion metallic tunnels. The underlying physical mechanisms are quantitatively explained through wave optics and optical surface transformation, indicating the size ratio between the incident and output surface governs both the steering angle and the compression ratio. Numerical simulations demonstrate its outstanding performance, achieving a maximum steering angle of 40° and a compression ratio of 0.4 across 3 to 12 GHz, with averaged energy transmittance above 80%. The experiments further validate its effectiveness by measuring the magnetic field distributions of the output beam at various frequencies. The excellent beam steering and compressing effects make the proposed device highly promising for next-generation multifunctional wave manipulation in advanced communication systems.

## I. INTRODUCTION

BEAM steering devices have broad applications in redirecting electromagnetic (EM) waves to enhance the scanning angle of a phased array antenna [1-4]. Current beam steering methods primarily involve: reflectarrays [5] and transmitarrays [6]; quasi-optical systems such as Rotman lenses [7] and Luneburg lenses [8]; metamaterials/metasurfaces with subwavelength artificial structures [9-12]. Despite the advancements of these approaches, several challenges persist in high-density scenarios. With the rapid development of IoT and 6G networks, the increasing number of mobile devices has led to a crowded communication environment. In such scenarios, a relatively wide half-power beamwidth (HPBW) may cause mutual interference of nodes, thereby constraining the maximum base station density and coverage efficiency. Furthermore, the propagating losses introduced by numerous circumstances (e.g. atmospheric losses, rain attenuation, high path loss, etc.) hinder the effective signal transmission, confining maximum communication and detection ranges [13]. To address these limitations, beam compressing devices have emerged to narrow the HPBW and enhance the energy density [14-16], enabling higher communication directivity and extending the effective distance. However, existing devices are not capable of simultaneously achieving beam steering and compressing, and cascading discrete beam steering devices and beam compressing devices would significantly increase the system's complexity and overall volume. Therefore, developing a simultaneous beam steering and compressing device (BSCD) would be greatly beneficial for the minimization and functional integration of wave manipulation systems.

In recent years, optic null medium (ONM) [17, 18], a highly anisotropic metamaterial designed via extreme coordinate transformation based on transformation optics [19, 20], has demonstrated remarkable success in regulating fundamental properties of EM waves due to its unique ability to directionally project the wave along its principal axis while maintaining the energy transmittance close to 100%, which is referred as to optical surface transformation (OST) theory [21, 22]. By designing proper shape of the incident and output surface, and establishing projection correspondence for the points between the incident and output surface through ONM's principal axis, numerous wave manipulation devices such as hyperlens [23], invisible cloak [24], and also beam steering devices [25] can be achieved. However, the operational bandwidth of previous ONM-based beam steering devices remains highly limited [25], and no study has yet demonstrated the simultaneous realization of beam steering and compressing using ONM. Therefore, it is necessary to design and build a broadband BSCD which is capable to simultaneously steer and compress the EM wave across a wide frequency range.

Inspired by recent works employing acoustic tunnels with

This work was supported in part by the National Natural Science Foundation of China (Nos. 12274317, and 12374277), the Natural Science Foundation of Shanxi Province (202303021211054). *(Corresponding author: Fei Sun.)*

Dongguo Zhang was with the Key Lab of Advanced Transducers and Intelligent Control System, Ministry of Education and Shanxi Province, College of Physics and Optoelectronics, Taiyuan University of Technology, Taiyuan, China. He is now with the Department of Mechanical Engineering, Korea Advanced Institute of Science and Technology, Daejeon, Republic of Korea (e-mail: zhangdongguo@kaist.ac.kr).

Fei Sun, Qin Liao and Yichao Liu are with the Key Lab of Advanced Transducers and Intelligent Control System, Ministry of Education and Shanxi Province, College of Physics and Optoelectronics, Taiyuan University of Technology, Taiyuan, China (e-mail: sunfei@tyut.edu.cn).

Donguk Nam is with the Department of Mechanical Engineering, Korea Advanced Institute of Science and Technology, Daejeon, Republic of Korea (e-mail: dwnam@kaist.ac.kr).

protrusion structures as the reduced acoustic null medium [26, 27], we design a novel ONM composed of multiple subwavelength protrusion metallic tunnels (SPMTs) with varied cross-sectional width and protrusion length to develop a broadband BSCD applicable to simultaneously steer and compress the TM-polarized EM waves ranging from 3 to 12 GHz with averaged energy transmittance above 90%. The operating mechanism is explained by wave optics and OST, which indicates that different steering angles and compression ratios can be achieved by designing the size ratio between the incident and output surfaces. The device's performance is validated through both numerical simulations and experimental measurements. The proposed BSCD provides new perspectives for multi-functional wave-manipulation approach, and it's highly applicable in various wave manipulation scenarios, especially for the high-density point-to-point communications and sensing systems.

## II. DESIGN METHODS

Fig. 1 (a) illustrates the functionality of the proposed BSCD, which takes a columnar form along the $z$-axis, with an isosceles trapezoid cross-section on the $xy$-plane. When the beam incident onto the incident surface at an incident angle $\alpha_i$, it is guided through the internal SPMTs toward the output surface and emits at an output angle $\alpha_o = \alpha_i + \Delta\alpha$, where $\Delta\alpha$ is the steering angle determined by both $\alpha_i$ and the geometrical parameters of the BSCD. At the same time, the output HPBW $w_o$ is compressed compared to the incident HPBW $w_i$, i.e. $w_o = C \cdot w_i$, where $C$ is the compression ratio ranging from 0 to 1. The BSCD's cross-sectional structure in the $xy$-plane is depicted in Fig. 1 (b). The inner SPMTs are symmetrically numbered from 1 to M, extending from the center toward both sides. The total length of the BSCD along the $x$-axis is $l$. A magnified view in Fig. 1 (b) highlights the protrusion structures of SPMTs, where the spacing between protrusions is $d$, and both the protrusions and the tunnels share a uniform thickness $t$. The protrusion length $p$ and the tunnel's cross-sectional width $h$, which are key in determining the effective relative refractive index $n_r$ of the SPMT, are designed to be much smaller than the operational wavelength to ensure only a single mode propagates along the principal axis of each SPMT [26].

To verify that SPMTs with different $p/w$ ratios correspond to different $n_r$, we conducted simulations of the magnetic field distributions inside the SPMTs, as shown in Fig. 1(c). The geometric parameters are set as $t = 1$ mm, $d = 5$ mm, $l = 200$ mm, The maximum and minimum width of the SPMT are set as 9 mm and 4 mm, respectively. The $p/h$ ratios of the three SPMTs are 0, 0.2, and 0.31, respectively. A frequency of $f = 6$ GHz is chosen for demonstration. It is obvious that when $p/w$ increases, the wavelength of the transmitted wave decreases correspondingly, verifying a larger $n_r$ is achieved. The quantitative relationship between $n_r$ and $p/h$ for different working frequencies is illustrated in Fig. 1 (d), indicating a positive correlation between $n_r$ and $p/h$. By setting proper $p/h$, the total optical path of each SPMT can be designed equally to ensure the BSCD only alters the wavefront's size while preserving its shape. Moreover, when $p/h < 0.25$, the SPMT is nearly dispersionless, which is highly advantageous for broadband wave manipulation. Fig. 1 (e) shows the energy transmittance $T$ of SPMT as a function of frequencies $f$ and $p/h$. $T$ is defined as $T = P_o / P_i$, where $P_o$ and $P_i$ are the powers of output and incident beams, respectively. Except at frequencies that significantly deviate from the Fabry–Pérot (F-P) resonance condition, $T$ remains above 0.8 for most cases, demonstrating its ability to serve as an ONM to achieve OST capable of directionally projecting the wave with minimal transmission loss.

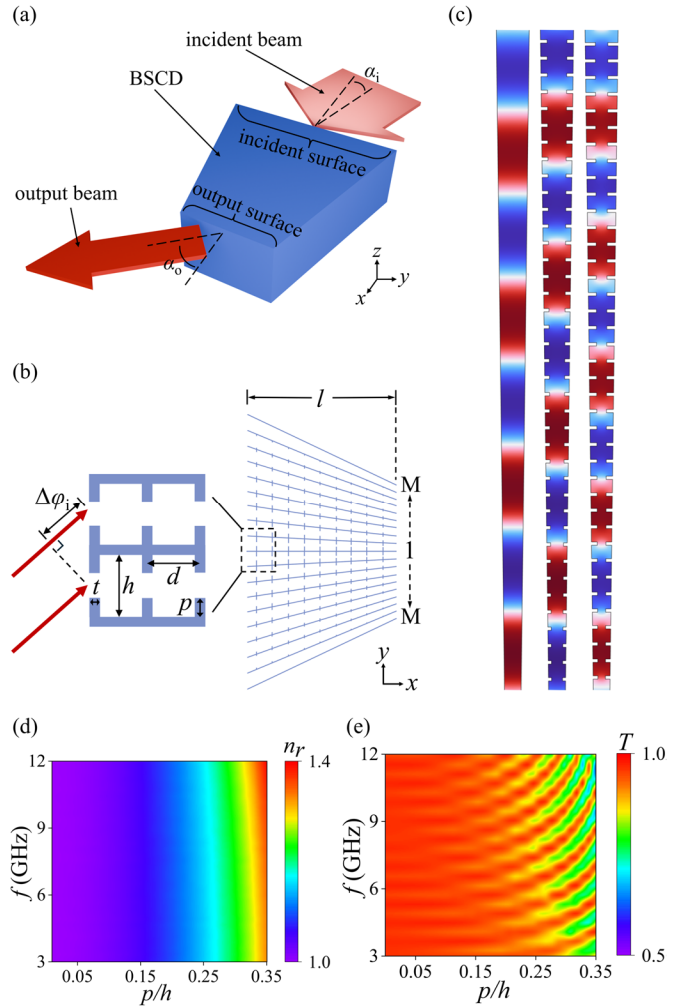

Fig. 1. (a) Schematic diagram of the simultaneous beam steering and compressing functionality of the BSCD. (b) Schematic diagram of the structure of BSCD in $xy$-plane – a trapezoidal profile with a total length $l$ composed of multiple SPMTs spaced at intervals of $h$. The protrusions with a length of $p$ and a thickness of $t$ are arranged inside the SPMTs at intervals of $d$. The symmetrically arranged SPMTs along the $x$-axis share the same sequence number 1 to M from the center to the sides. (c) Simulated magnetic field distributions that vary with the ratio $p/h = 0$, 0.2, and 0.31, respectively. (d) Schematic diagram of effective relative refractive index varies with frequency and $p/h$. (e) Energy transmittance $T$ under normal incidence as a function of $p/h$ and frequency $f$ of incident waves.

## III. OPERATING MECHANISM

In this chapter, we explain the steering and compressing effects of the BSCD using wave optics and OST, respectively, and present the relationships between the steering angle, compression ratio and the geometric parameters of the BSCD. The methodology for designing the structural parameters to achieve targeted index of the beam steering and compressing is provided as well.

The beam steering mechanism of the proposed BSCD is explained as follow: Considering a plane wave incident on the BSCD at $\alpha_i$, as shown by the red arrows in the magnified illustration of Fig. 1 (b), the phase difference $\Delta\varphi_i$ between two adjacent SPMTs at the incident surface is given by:

$$\Delta\varphi_i = (h_i + t) \cdot \sin\alpha_i \cdot 2\pi / \lambda, \quad (1)$$

where $h_i$ is the tunnel's cross-sectional width at the incident surface, $\lambda$ is the wavelength of the incident wave. Similarly, the phase difference $\Delta\varphi_o$ at the output surface can be described as:

$$\Delta\varphi_o = (h_o + t) \cdot \sin\alpha_o \cdot 2\pi / \lambda. \quad (2)$$

Considering that the total optical paths of the SPMTs are designed to be equal, $\Delta\varphi_i$ should be equal to $\Delta\varphi_o$ as well:

$$\Delta\varphi_i = \Delta\varphi_o. \quad (3)$$

Defining $\gamma = (h_o + t) / (h_i + t)$, and combining the above equations, the relationship between the steering angle and the incident angle can be derived:

$$\Delta\alpha = \arcsin(\gamma^{-1}\sin\alpha_i) - \alpha_i. \quad (4)$$

The critical angle $\alpha_c$ (i.e., the smallest incident angle that yields total reflection) can be expressed as:

$$\alpha_c = \arcsin\gamma. \quad (5)$$

The beam compressing mechanism can be explained by the OST theory [21, 22], which does not need to consider any mathematical calculation during the designing process, as the wavefront is directly determined by the shapes of the incident and output surface. In the proposed BSCD, both surfaces are planar, so the wavefront shape remains unchanged while only its size is altered. Despite the diffraction effects, the compression ratio $C$ can be simply predicted as:

$$C = \gamma. \quad (6)$$

To conclude, the parameter $\gamma$ directly influences both the steering angle and compression ratio. When the incident angle remains fixed, a smaller $\gamma$ leads to a larger steering angle and a smaller compression ratio, while the numerical aperture, which is directly decided by the critical angle, will correspondingly decrease. Besides, the steering angle increases with the incident angle as well.

## IV. NUMERICAL SIMULATIONS

To demonstrate the BSCD's simultaneous beam steering and compressing effect, numerical simulations are conducted for EM Gaussian beams incident at various angles. The BSCD parameters employed in the simulations are designed as follows: $M = 18$, $t = 1$ mm, $d = 5$ mm, $l = 200$ mm, $h_i = 9$ mm, $h_o = 4$ mm, with the $p/h$ ratios for each SPMT provided in Table I. All simulations are conducted by COMSOL Multiphysics 5.6 in 2D

TABLE I
$p/h$ OF EACH SPMT IN THE SIMULATIONS

| sequence number | p/h | sequence number | p/h |
|---|---|---|---|
| 1 | 0.213 | 10 | 0.187 |
| 2 | 0.212 | 11 | 0.178 |
| 3 | 0.211 | 12 | 0.167 |
| 4 | 0.209 | 13 | 0.161 |
| 5 | 0.208 | 14 | 0.156 |
| 6 | 0.206 | 15 | 0.128 |
| 7 | 0.200 | 16 | 0.112 |
| 8 | 0.194 | 17 | 0.111 |
| 9 | 0.192 | 18 | 0.100 |

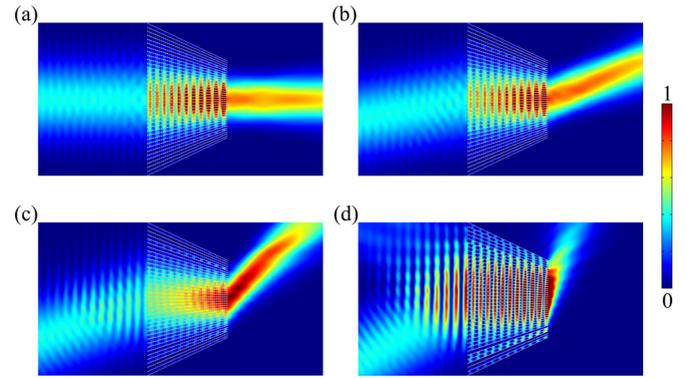

Fig. 2. Simulation results showing the normalized intensity distribution of the magnetic field's z-component for Gaussian beams with frequency $f = 7.18$ GHz incident on the BSCD with incident angels of (a) 0°, (b) 10°, (c) 20° and (d) 30°.

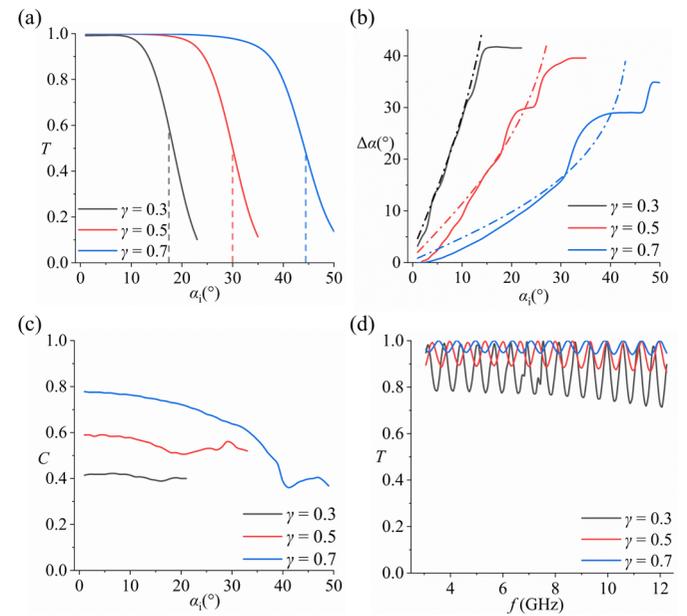

Fig. 3. Simulated results for three different $\gamma$ values showing the relationship between: (a) the energy transmittance $T$ and the incident angle $\alpha_i$ (The dashed lines indicate the critical angles derived from Eq. 5.); (b) the steering angle $\Delta\alpha$ and $\alpha_i$ (The dashed lines are theoretical expectations according to Eq. 4.); (c) the compression ratio $C$ and $\alpha_i$; (d) $T$ and operating frequency $f$.

cases, utilizing the wave optics module with a steady-state solver to simulate EM waves. Free triangle meshing is used,

where the maximum grid is one-tenth of the minimum working wavelength ($\lambda_{min}$ = 21.5 mm) to ensure accuracy of simulations.

Fig. 2 (a) to (d) show the absolute value of the simulated normalized $z$-component of the magnetic field distributions, $|H_z|$, for a wave with frequency $f$ = 7.18 GHz incident at angles of 0°, 10°, 20°, and 30°, respectively. In each case, the beams are guided along the SPMTs' principal axis, and the output HPBWs are obviously compressed. For oblique incident scenarios, the output angle $\alpha_o$ is larger than the incident angle $\alpha_i$, leading to an increasing steering angle $\Delta\alpha$ as $\alpha_i$ grows. The steering angles for these four cases are 0°, 8.8°, 27.7° and 38.6°, respectively.

To investigate how different structural parameters influence the BSCD's beam steering and compression performance, simulations are performed for BSCDs with $\gamma$ values of 0.7, 0.5 and 0.3 at 7.18 GHz. Fig. 3 (a) shows the energy transmittance $T$ as a function of the incident angle $\alpha_i$. When $\alpha_i$ is less than 10°, $T$ approaches to 1. As $\alpha_i$ increases, $T$ begins to decrease correspondingly. Interestingly, $T$ still remains a relatively high value of about 0.5 when $\alpha_i$ approaches the critical angle $\alpha_c$ indicated by dashed lines, which are theoretically derived from Eq. (5). Fig. 3 (b) indicates the relationship between $\alpha_i$ and $\Delta\alpha$, where the solid and dashed lines correspond to the simulations and the solutions of Eq. 4, respectively, both illustrating a positive correlation between $\alpha_i$ and $\Delta\alpha$. For the same $\alpha_i$, a smaller $\gamma$ produces a larger $\Delta\alpha$, consistent with Eq. (4). Fig. 3 (c) illustrates the compression ratio $C$ over a range of $\alpha_i$ values, where $C$ does not change significantly within a wide range of $\alpha_i$. Furthermore, for a smaller $\gamma$, $C$ is smaller correspondingly, which is consistent with the prediction of Eq. (6). Fig. 3 (d) depicts the energy transmittance $T$ for different frequencies $f$ under normal incidence. In most cases, $T$ exceeds 0.8, and the averaged $T$ exceeds 0.9 for all the three $\gamma$ values, confirming the BSCD's broadband working capability from 3 to 12 GHz. Moreover, $T$ reaches its maximum when the frequency satisfies F-P resonance condition.

To conclude, the $\gamma$ value plays a key role in influencing both the steering and compressing effects when $\alpha_i$ and $f$ are fixed, as discussed in the previous chapter. A smaller $\gamma$ corresponds to a larger $\Delta\alpha$ and a smaller $C$, resulting in enhanced steering and compressing performance. However, reducing $\gamma$ can negatively impact energy transmittance $T$—not only because it leads to a smaller critical angle $\alpha_c$, but also because $T$ under normal incidence tends to be lower across most frequencies. Besides, a larger $\alpha_i$ negatively affects the transmittance $T$, but results in a greater $\Delta\alpha$. The influence of frequency $f$ on $T$ is primarily determined by F-P resonance, and $T$ remains above 70% across the 3–12 GHz range even when the F-P resonance condition is not satisfied.

## V. Experimental Measurements

To verify the simultaneous beam steering and compression effects, a BSCD sample is fabricated and experimentally tested, with the measurement setup illustrated in Fig. 4(a). The parameters of the fabricated BSCD sample are set as follows: M = 11, $h_i$ = 14 mm, $h_o$ = 6 mm, $l$ = 200 mm, $t$ = 1 mm, $d$ = 5 mm, $p/h$ for each SPMT is listed in Table Ⅱ. The EM signal is generated at port 1 of a vector network analyzer (VNA). It is then connected via a horn antenna to generate the incident EM

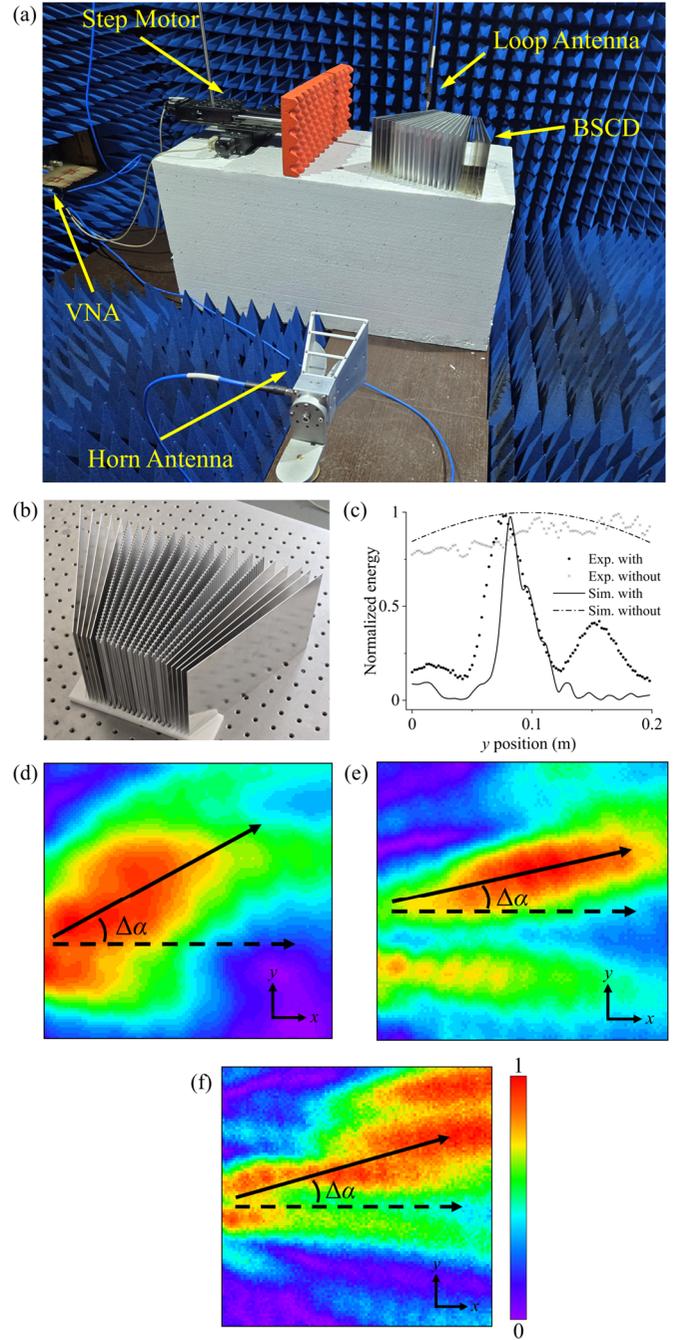

Fig. 4. (a) Experimental setup: The BSCD sample is mounted on a foam sheet platform approximately 1 meter above the ground. A horn antenna, connected to VNA port 1, generates EM waves and is positioned 1 m in front of the BSCD. VNA port 2, connected to a loop antenna, is used to measure the $H_z$ field distribution behind the BSCD; the loop antenna is controlled by a step motor. (b) Photograph of the fabricated BSCD sample, comprising a series of SPMTs produced by CNC machining and fixed by a 3D-printed resin base. (c) Simulated and measured results for the normalized energy distribution of the output beam along the $y$-axis at 0.1 meters behind the output surface, compared with the reference case (without the BSCD). The frequency is set as 9 GHz. The dots and stars represent the experimental results with and without the BSCD, respectively, while the solid and dashed lines represent the corresponding simulated results. (d-f) Measured normalized $|H_z|$ field distribution of the output beam at an incident angle of 20° and frequencies of 7 GHz, 9 GHz, and 11 GHz, respectively.

wave, which is placed at 1 meter in front of the BSCD sample. A loop antenna is placed behind the output surface of the BSCD

TABLE II
$p/h$ OF EACH SPMT IN THE FABRICATED BSCD SAMPLE

| sequence number | p/h | sequence number | p/h |
| --- | --- | --- | --- |
| 1 | 0.152 | 7 | 0.133 |
| 2 | 0.152 | 8 | 0 |
| 3 | 0.152 | 9 | 0 |
| 4 | 0.152 | 10 | 0 |
| 5 | 0.133 | 11 | 0 |
| 6 | 0.133 | 12 | 0 |

sample and controlled by a step motor to measure the magnetic field's z-component $H_z$ of the output beam. Subsequently, the loop antenna is connected via a coaxial cable to VNA port 2 for measuring the $S_{21}$ parameters. The BSCD sample, the horn antenna and the loop antenna are on the same horizontal plane, and the entire system is placed in the microwave darkroom to avoid background noise. When the horn antenna remains fixed, varying the BSCD's orientation could provide different incident angles $α_i$. Under oblique incidence, a deviation of the output beam's direction from the horn antenna's orientation could validate the beam steering effect. Fig. 4 (b) shows the fabricated BSCD sample, composing of multiple SPMTs with varying cross-section width and protrusion length, fabricated from aluminum alloy 1060 via computerized numerical control (CNC) machining. A 3D-printed resin base ensures each SPMT remains in its designated position, preserving the intended shape.

Fig. 4 (c) plots the measured and simulated normalized energy distribution along the y-axis at 0.1 meters behind the output surface when the BSCD sample is rotated by 20° (i.e. the incident angle is 20°), alongside a reference measurement without the BSCD at the same position. The frequency of the incident beam is set as 9 GHz. The dots and stars indicate the measured results with and without BSCD, respectively, while the solid and dashed lines show the corresponding simulated results, demonstrating great agreement with the experiments. Compared to the nearly plane-wave profile of the reference beam, the output beam's HPBW = 41 mm is distinctly compressed. Fig. 4 (d-f) illustrates the measured normalized $|H_z|$ distributions of the output beams at an incident angle of 20° and frequencies of 7 GHz, 9 GHz, and 11 GHz, respectively, where their propagation directions (indicated by the solid arrows) significantly deviate from that of the incident beams (indicated by the dashed arrows). Therefore, the simultaneous beam steering and compressing effect of the proposed BSCD is validated experimentally.

## VI. CONCLUSION

In this work, a novel BSCD is designed using wave optics and OST, and implemented via SPMTs. Numerical simulations are conducted to investigate its beam steering and compressing performance for the TM-polarized waves across 3 to 12 GHz with various size ratios between the incident and output surface. Results show that a smaller size ratio plays a key role in increasing the steering angle and decreasing the compression ratio, with only a slight reduction in energy transmittance. We also fabricate a sample to experimentally validate its performance at different frequencies by measuring the magnetic field distributions of the output beams. This study introduces the first wave manipulation device capable of concurrently steering a beam and compressing its HPBW. The SPMT-based BSCD unveils novel approaches to multifunctional wavefront control, with promising applications in various scenarios that require complex manipulations of EM waves.